\RequirePackage{fix-cm}
\documentclass[10pt,onecolumn]{IEEEtran}

\newtheorem{Propo}{\bf Proposition}[section]
\usepackage{epsfig}
\usepackage{cite}
\usepackage[english]{babel}
\usepackage{amsmath}
\usepackage{amsfonts}
\usepackage{amssymb}
\usepackage{mathrsfs}
\usepackage[dvipsnames]{xcolor}
\usepackage[breaklinks=false,colorlinks]{hyperref}
\begin{document}

\title{Data Detection in Single User Massive MIMO Using Re-Transmissions}

\author{
K. Vasudevan, K. Madhu, and Shivani Singh

\thanks{
The authors are with the Department of Electrical Engineering, Indian
Institute of Technology Kanpur, 208016, India.
email:\{vasu, kmadhu, shivanis\}@iitk.ac.in
}}




\maketitle
\begin{abstract}
Single user massive multiple input multiple output (MIMO) can be used to
increase the spectral efficiency, since the data is transmitted simultaneously
from a large number of antennas located at both the base station and mobile.
It is feasible to have a large number of antennas in the mobile, in the
millimeter wave frequencies.
However, the major drawback of single user massive MIMO is
the high complexity of data recovery at the receiver. In this work, we propose
a low complexity method of data detection with the help of re-transmissions. A
turbo code is used to improve the bit-error-rate (BER). Simulation
results indicate significant improvement in BER with just two re-transmissions
as compared to the single transmission case. We also show that the minimum
average SNR per bit required for error free propagation over a massive MIMO
channel with re-transmissions is identical to that of the additive white
Gaussian noise (AWGN) channel, which is equal to $-1.6$ dB.
\end{abstract}
\section{Introduction}
The main idea behind single user massive multiple input multiple output (MIMO)
\cite{6770094,6798744,738086,6375940,6736761,7047295,7064850} is to increase
the bit rate between the transmitter and receiver over a wireless
channel. This is made possible by sending the bits or symbols (groups of bits)
simultaneously from a large number of transmit antennas. The signal at each
receive antenna is a linear combination of the bits or symbols sent from all
the transmit antennas plus additive white Gaussian noise (AWGN). We assume that
the carrier frequency offset is absent or has been accurately estimated
and canceled with the help of training symbols (preamble)
\cite{6663392,Vasudevan2015,DBLP:journals/corr/Vasudevan15a,Vasu_ICWMC2016,Vasu_Adv_Tele_2017}. The task of the
receiver is to estimate the transmitted bits or symbols, from the signals in
all the receive antennas. Note that it is possible to have a large number of
antennas in both the base station and the mobile, in millimeter wave
frequencies \cite{5783993,6515173,6736750,6824752,6882985,7032050,7574380,7601145,7841750}, due to the small size of the antennas.







If both the transmitter and receiver have $N$ antennas and the symbols are
drawn from an $M$-ary constellation, the complexity of the maximum likelihood
(ML) detector would be $M^N$, since it exhaustively searches all possible
symbol combinations. Clearly, the ML detector is impractical. On the other
hand, the ``zero-forcing'' solution is to multiply the received signal
vector by the inverse of the channel matrix, which eliminates the interference
from the other symbols. However, the computational complexity of
inversion of the $N\times N$ channel matrix, for large values of $N$, does
not make this approach attractive. Moreover, when the noise vector is
multiplied by the inverse of the channel matrix, it usually results in
noise enhancement, leading to poor bit-error-rate (BER) performance.

In \cite{8109972}, a split pre-conditioned conjugate gradient method for
data detection in massive MIMO is proposed. A low-complexity soft-output
data detection scheme based on Jacobi method is presented in \cite{7996693}.
Near optimal data detection based on steepest descent and Jacobi method is
presented in \cite{7342925}. Matrix inversion based on Newton iteration for
large antenna arrays is given in \cite{7370771}. Subspace methods of data
detection in MIMO are presented in \cite{6139804,7925914}. Data detection in
large scale MIMO systems using successive interference cancellation (SIC) is
given in \cite{7791076}. MIMO data detection in the presence of phase noise is
given in \cite{7447017}. Detection of LDPC coded symbols in MIMO systems
is discussed in \cite{8050314}. Decoding of convolutional codes in MIMO systems
is presented in \cite{6138882}. Decoding of polar codes in MIMO systems is
given in \cite{7527195}. Sphere decoding procedures for the detection of
symbols in MIMO systems are discussed in \cite{1603705,6271905,7801954}. Large
scale MIMO detection algorithms are presented in \cite{6777306}. Multiuser
detection in massive MIMO with power efficient low resolution ADCs is given in
\cite{6987288}. Joint ML detection and channel estimation
in multiuser massive MIMO is presented in \cite{7439790}. Detection of turbo
coded offset QPSK in the presence of frequency and clock offsets and AWGN
is presented in \cite{Vasu_SIVP10,DBLP:journals/corr/Vasudevan15}.
Channel estimation in large antenna systems is given in
\cite{6808541,7073756,7160668}. Channel-aware data fusion for massive MIMO,
in the context of wireless sensor networks (WSNs) is proposed in \cite{6971234}.

In all the papers in the literature, on the topic of data detection in massive
MIMO, the main lacuna has been in the definition of (or rather the lack of it)
the signal-to-noise ratio (SNR). In fact, even the operating SNR of a mobile
phone is not known \cite{Vasudevan2015,DBLP:journals/corr/Vasudevan15a,Vasu_ICWMC2016,Vasu_Adv_Tele_2017,Vasu_MAAZE_Edit16}. It may be
noted that the mobile phones indicate a typical signal strength of $-100$ dBm
($10^{-10}$ milliwatt). However, this is not the SNR. In this work, we use the
SNR per bit as the performance measure, since there is a lower bound on the
SNR per bit for error-free transmission over any type of channel, which is
$-1.6$ dB \cite{Vasu_ICWMC2016,Vasu_Adv_Tele_2017}. The so-called ``capacity''
of MIMO channels has been derived earlier in
\cite{Telatar:1999:1124-318X:585,1203154,4907090,6809207}. However, the
channel capacity is derived differently in
\cite{Vasu_ICWMC2016,Vasu_Adv_Tele_2017}, and in this work.
Therefore, the question naturally arises: are the present-day wireless
telecommunication systems operating anywhere near the channel capacity? This
question assumes significance in the context of 5G wireless communications
where not only humans, but also machines and devices would be connected to
the internet to form the Internet of Things (IoT) \cite{7414384}. Hence, in
order to minimize the global energy consumption due to IoT, it is necessary
for each device to operate as close to the minimum average SNR per bit for
error-free transmission, as possible
\cite{Vasudevan2015,DBLP:journals/corr/Vasudevan15a,Vasu_ICWMC2016,Vasu_Adv_Tele_2017,Vasu_MAAZE_Edit16,6736745}.
Finally, one might ask the question: is it not possible to increase the
bit-rate by increasing the size of the constellation, and using just one
transmit and receive antenna? The answer is: increasing the size of the
constellation increases the peak-to-average power ratio (PAPR), which
poses a problem for the radio frequency (RF)
front end amplifiers, in terms of the dynamic range. In other words, a
large PAPR requires a large dynamic range, which translates to low power
efficiency, for the RF amplifiers \cite{halkias2001integrated}.
 
In this work, we re-transmit a symbol $N_{rt}$ times and then take the
average, which results in a lower interference power compared to the single
transmission case. Perfect channel state information (CSI) is assumed. The
BER is improved with the help of a turbo code.
This paper is organized as follows. Section~\ref{Sec:System_Model}  describes
the system model. The receiver design is presented in
Section~\ref{Sec:Receiver}. The bit-error-rate (BER) results from computer
simulations are given in Section~\ref{Sec:Sim_Results}. Finally,
Section~\ref{Sec:Conclude} presents the conclusions and future work.


\section{System Model}
\label{Sec:System_Model}
\begin{figure}[tbh]
\begin{center}
\input{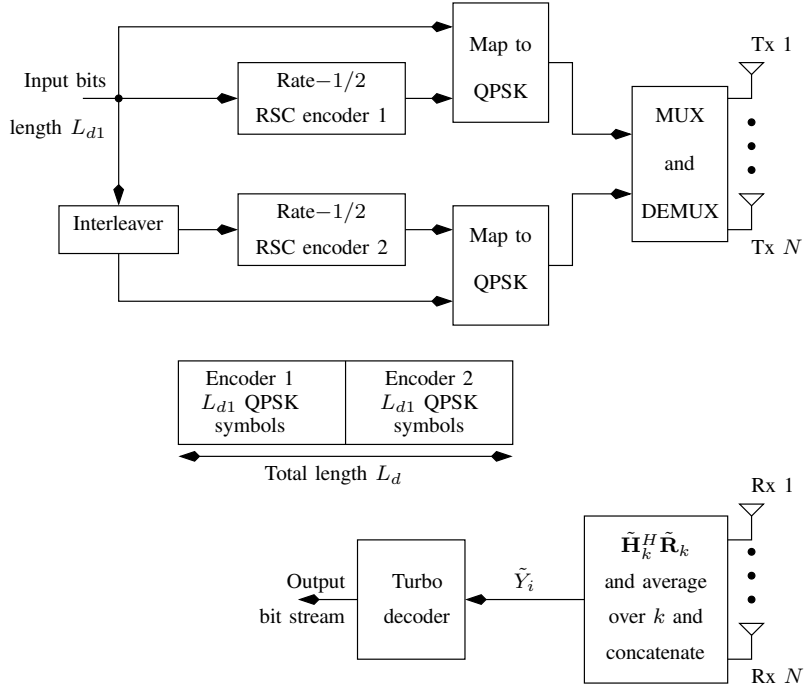}
\caption{System model.}
\label{Fig:System_Model}
\end{center}
\end{figure}
Consider the system model in Figure~\ref{Fig:System_Model}. The data bits are
organized into frames of length $L_{d1}$ bits. The recursive systematic
convolutional (RSC) encoders 1 and 2 encode the data bits into
Quadrature Phase Shift Keyed (QPSK) symbols having a total length of $L_d$.
We assume a MIMO system with  $N$ transmit and $N$ receive antennas. We also
assume that $L_d/N$ is an integer, where $L_{d}=2L_{d1}$ as shown
in Figure~\ref{Fig:System_Model}. The $L_d$ QPSK symbols are transmitted,
$N$ symbols at a time, from the $N$ transmit antennas.

The received signal in the $k^{th}$ ($0\leq k\leq N_{rt}-1$,
$k$ is an integer), re-transmission  is given by
\cite{Vasu_ICWMC2016,Vasu_Adv_Tele_2017}
\begin{equation}
\label{Eq:Massive_MIMO_Eq1}
\tilde{\mathbf{R}}_k = \tilde{\mathbf{H}}_k
                       \mathbf{S} +
                       \tilde{\mathbf{W}}_k
\end{equation}
where $\tilde{\mathbf{R}}_k\in \mathbb{C}^{N\times 1}$ is the received
vector, $\tilde{\mathbf{H}}_k\in \mathbb{C}^{N\times N}$ is the channel
matrix and $\tilde{\mathbf{W}}_k\in \mathbb{C}^{N\times 1}$ is the additive 
white Gaussian noise (AWGN) vector. The transmitted symbol vector is
$\mathbf{S}\in \mathbb{C}^{N\times 1}$, whose elements are drawn from an
$M$-ary constellation. Boldface letters denote vectors or matrices. Complex
quantities are denoted by a  tilde. However tilde is not used for complex
symbols $\mathbf{S}$. The elements of $\tilde{\mathbf{H}}_{k}$ are
statistically independent with zero  mean and variance per dimension equal to
$\sigma_{H}^{2}$, that is 
\begin{equation}
\label{Eq:Massive_MIMO_Eq2}
\frac{1}{2}
 E
\left[
\left|
\tilde{H}_{k,\, i,\, j}
\right|^{2}
\right] = \sigma_{H}^2
\end{equation}
where $E[\cdot]$ denotes the expectation operator \cite{Haykin83,Vasu_Book10},
$\tilde{H}_{k,\, i,\, j}$ denotes the element in the $i^{th}$ row and $j^{th}$
column of $\tilde{\mathbf{H}}_k$. Similarly, the elements of
$\tilde{\mathbf{W}}_k$ are statistically independent with zero mean and 
variance per dimension equal to $\sigma_W^2$, that is 
\begin{equation}
\label{Eq:Massive_MIMO_Eq3}
\frac{1}{2}
 E
\left[
\left|
\tilde{W}_{k,\, i}
\right|^{2}
\right] = \sigma_W^2
\end{equation}
where $\tilde{W}_{k,\, i}$ denotes the element in the $i^{th}$ row of 
$\tilde{\mathbf{W}}_{k}$. The real and imaginary parts of
$\tilde{H}_{k,\, i,\, j}$ and $\tilde{W}_{k,\, i}$ are also assumed to be
independent. The channel and noise are assumed to be independent across
re-transmissions, that is
\begin{equation}
\label{Eq:Massive_MIMO_Eq4}
\begin{aligned}
\frac{1}{2}
 E
\left[
\tilde{\mathbf{H}}_k
\tilde{\mathbf{H}}_l^H
\right] & =  N
            \sigma_H^2
            \delta_K(k-l)
            \mathbf{I}_N   \\
\frac{1}{2}
 E
\left[
\tilde{\mathbf{W}}_k
\tilde{\mathbf{W}}_l^H
\right] & = \sigma_W^2
            \delta_K(k-l)
            \mathbf{I}_N   \\
\end{aligned}
\end{equation}
where the superscript $(\cdot)^H$ denotes Hermitian (conjugate transpose of
a matrix), $\mathbf{I}_N$ is an $N\times N$ identity matrix and $\delta_K(m)$
($m$ is an integer) is the Kronecker delta function defined by
\begin{equation}
\label{Eq:Massive_MIMO_Eq4_1}
\delta_K(m) =
\left
\{
\begin{array}{ll}
1 & \mbox{for $m=0$}\\
0 & \mbox{otherwise.}
\end{array}
\right.
\end{equation}
The receiver is  assumed to have perfect knowledge of $\tilde{\mathbf{H}}_k$. 
\section{Receiver}
\label{Sec:Receiver}
In this section, we describe the procedure for detecting $\mathbf{S}$ given
the received signal $\tilde{\mathbf{R}}_k$ in (\ref{Eq:Massive_MIMO_Eq1}). 
Consider
\begin{eqnarray}
\label{Eq:Massive_MIMO_Eq5}
\tilde{\mathbf{Y}}_k & = & \tilde{\mathbf{H}}_k^H
                           \tilde{\mathbf{R}}_k  \nonumber  \\ 
                     & = & \tilde{\mathbf{F}}_k\mathbf{S}+
                           \tilde{\mathbf{V}}_k
\end{eqnarray}
where
\begin{equation}
\label{Eq:Massive_MIMO_Eq6}
\begin{aligned}
\tilde{\mathbf{F}}_k & = \tilde{\mathbf{H}}_k^H\tilde{\mathbf{H}}_k\\ 
\tilde{\mathbf{V}}_k & = \tilde{\mathbf{H}}_k^{H}\tilde{\mathbf{W}}_k.
\end{aligned}
\end{equation}
Observe that similar to (\ref{Eq:Massive_MIMO_Eq4}) we have
\begin{equation}
\label{Eq:Massive_MIMO_Eq6_1}
\frac{1}{2}
 E
\left[
\tilde{\mathbf{H}}_k^H
\tilde{\mathbf{H}}_l
\right]   =  N
            \sigma_H^2
            \delta_K(k-l)
            \mathbf{I}_N.
\end{equation}
However
\begin{equation}
\label{Eq:Massive_MIMO_Eq6_2}
\frac{1}{2}
\tilde{\mathbf{H}}_k^H
\tilde{\mathbf{H}}_l
            \neq
             N
            \sigma_H^2
            \delta_K(k-l)
            \mathbf{I}_N.
\end{equation}
The main aim of this work is to replace the expectation operator in
(\ref{Eq:Massive_MIMO_Eq6_1}) by time-averaging, in the form of
re-transmissions, so that the right-hand-side of (\ref{Eq:Massive_MIMO_Eq6_1})
is approximately satisfied.

Now the $i^{th}$ element of $\tilde{\mathbf{Y}}_k$
in (\ref{Eq:Massive_MIMO_Eq5}) is 
\begin{equation}
\label{Eq:Massive_MIMO_Eq7}
\tilde{Y}_{k,\, i} = \tilde{F}_{k,\, i,\, i} S_i +
                     \tilde{I}_{k,\, i} +
                     \tilde{V}_{k,\, i}
                     \quad \mbox{for $0\leq i\leq N-1$}
\end{equation}
where
\begin{equation}
\label{Eq:Massive_MIMO_Eq8}
\begin{aligned}
\tilde{F}_{k,\, i,\, i} & = \sum_{j=1}^{N}
                            \left|
                            \tilde{H}_{k,\, j,\, i}
                            \right|^2                                    \\ 
\tilde{I}_{k,\, i}      & = \sum_{j=1,\, j\neq i}^{N}
                            \tilde{F}_{k,\, i,\, j} S_j                  \\ 
\tilde{F}_{k,\, i,\, j} & = \sum_{l=1}^{N}
                            \tilde{H}_{k,\, l,\, i}^*
                            \tilde{H}_{k,\, l,\, j}                      \\ 
\tilde{V}_{k,\, i}      & = \sum_{j=1}^{N}
                            \tilde{H}_{k,\, j,\, i}^*
                            \tilde{W}_{k,\, j}
\end{aligned}
\end{equation}
where it is understood that $\tilde F_{k,\, i,\, i}$ is real-valued.
Note that for large values of $N$, $\tilde I_{k,\, i}$ and $\tilde V_{k,\, i}$
are Gaussian distributed due to the central limit theorem \cite{Haykin83}.
Moreover, since $S_i$ and $\tilde W_{k,\, i}$ are independent,
$\tilde I_{k,\, i}$ and $\tilde V_{k,\, i}$ are uncorrelated, that is
\begin{eqnarray}
\label{Eq:Massive_MIMO_Eq8_1}
 E
\left[
\tilde I_{k,\, i}
\tilde V_{k,\, i}^*
\right] & = &  E
              \left[
              \left(
              \sum_{j=1,\,  j\ne i}^{N}
              \tilde{F}_{k,\, i,\, j} S_j
              \right)
              \left(
              \sum_{l=1}^{N}
              \tilde{H}_{k,\, l,\, i}^*
              \tilde{W}_{k,\, l}
              \right)^*
              \right]                              \nonumber  \\
        & = & \sum_{j=1,\,  j\ne i}^{N}
              \sum_{l=1}^{N}
               E
              \left[
              \tilde{F}_{k,\, i,\, j}
              \tilde{H}_{k,\, l,\, i}
              \right]
               E
              \left[
               S_j
              \right]
               E
              \left[
              \tilde{W}_{k,\, l}^*
              \right]                              \nonumber  \\
        & = &  0.
\end{eqnarray}
Let
\begin{equation}
\label{Eq:Massive_MIMO_Eq9}
\tilde{U}_{k,\, i}' = \tilde{I}_{k,\, i} +
                      \tilde{V}_{k,\, i}
\end{equation}
where $\tilde I_{k,\, i}$ denotes the interference and $\tilde V_{k,\, i}$
denotes the noise term. From (\ref{Eq:Massive_MIMO_Eq8_1}) we have
\begin{eqnarray}
\label{Eq:Massive_MIMO_Eq10}
 E
\left[
\left|
\tilde{U'}_{k,\, i}
\right|^2
\right] & = &  E
              \left[
              \left|
              \tilde{I}_{k,\, i}
              \right|^2
              \right] +
               E
              \left[
              \left|
              \tilde{V}_{k,\, i}
              \right|^2
              \right ]                        \nonumber  \\ 
        & \stackrel{\Delta}{=}
            & \sigma_{U'}^2.
\end{eqnarray}
The noise power is 
\begin{eqnarray}
\label{Eq:Massive_MIMO_Eq11}
 E
\left[
\left|
\tilde{V}_{k,\, i}
\right|^2
\right] & = &  E
              \left[
              \left(
              \sum_{m=1}^{N}
              \tilde{H}_{k,\, m,\, i}^*
              \tilde{W}_{k,\, m}
              \right)
              \left(
              \sum_{n=1}^{N}
              \tilde{H}_{k,\, n,\, i}
              \tilde{W}_{k,\, n}^*
              \right)
              \right]                                   \nonumber  \\
        & = & \sum_{m=1}^{N}
              \sum_{n=1}^{N}
               E
              \left[
              \tilde{H}_{k,\, n,\, i}
              \tilde{H}_{k,\, m,\, i}^*
              \right]
               E
              \left[
              \tilde{W}_{k,\, m}
              \tilde{W}_{k,\, n}^*
              \right]                                   \nonumber  \\
        & = & \sum_{m=1}^{N}
              \sum_{n=1}^{N}
               2
              \sigma_{H}^2
              \delta_K(m-n)
               2
              \sigma_{W}^2
              \delta_K(m-n)                             \nonumber  \\
        & = &  4N
              \sigma_{H}^2
              \sigma_{W}^2
\end{eqnarray}
where we have used the sifting property of the Kronecker delta function.
The interference power is 
\begin{eqnarray}
\label{Eq:Massive_MIMO_Eq12}
 E
\left[
\left|
\tilde{I}_{k,\, i}
\right|^2
\right] & = &  E
              \left[
              \left(
              \sum_{m=1,\, m\neq i}^{N}
              \tilde{F}_{k,\, i,\, m} S_m
              \right)
              \left(
              \sum_{n=1,\, n\neq i}^{N}
              \tilde{F}_{k,\, i,\, n}^* S_n^*
              \right)
              \right]                                       \nonumber  \\
        & = & \sum_{m=1,\, m\ne i}^{N}
              \sum_{n=1,\, n\neq i}^{N}
               E
              \left[
              \tilde{F}_{k,\, i,\, m}
              \tilde{F}_{k,\, i,\, n}^*
              \right]
               E
              \left[
               S_m S_n^*
              \right]                                       \nonumber  \\
        & = & \sum_{m=1,\, m\ne i}^{N}
              \sum_{n=1,\, n\neq i}^{N}
               E
              \left[
              \tilde{F}_{k,\, i,\, m}
              \tilde{F}_{k,\, i,\, n}^*
              \right]
               P_{\mathrm{av}}
              \delta_K(m-n)                                 \nonumber  \\
        & = & \sum_{m=1,\, m\ne i}^{N}
               E
              \left[
              \left|
              \tilde{F}_{k,\, i,\, m}
              \right|^2
              \right]
               P_{\mathrm{av}}                              \nonumber  \\
        & = &  8 N (N-1) \sigma_H^4
\end{eqnarray}
where
\begin{eqnarray}
\label{Eq:Massive_MIMO_Eq13}
 E
\left[
\left|
 S_m
\right|^2
\right] & = &  P_{\mathrm{av}}                         \nonumber  \\
        & = &  2
\end{eqnarray}
and
\begin{eqnarray}
\label{Eq:Massive_MIMO_Eq13_1}
 E
\left[
\left|
\tilde{F}_{k,\, i,\, m}
\right|^2
\right] & = &  E
              \left[
              \left(
              \sum_{j=1}^{N}
              \tilde H_{k,\, j,\, i}^*
              \tilde H_{k,\, j,\, m}
              \right)
              \left(
              \sum_{l=1}^{N}
              \tilde H_{k,\, l,\, i}
              \tilde H_{k,\, l,\, m}^*
              \right)
              \right]                                  \nonumber  \\
        & = & \sum_{j=1}^{N}
              \sum_{l=1}^{N}
               E
              \left[
              \tilde H_{k,\, l,\, i}
              \tilde H_{k,\, j,\, i}^*
              \tilde H_{k,\, j,\, m}
              \tilde H_{k,\, l,\, m}^*
              \right]                                  \nonumber  \\
        & = & \sum_{j=1}^{N}
              \sum_{l=1}^{N}
               E
              \left[
              \tilde H_{k,\, l,\, i}
              \tilde H_{k,\, j,\, i}^*
              \right]
               E
              \left[
              \tilde H_{k,\, j,\, m}
              \tilde H_{k,\, l,\, m}^*
              \right]                                  \nonumber  \\
        & = & \sum_{j=1}^{N}
              \sum_{l=1}^{N}
               2\sigma^2_H
              \delta_K(l-j)
               2\sigma^2_H
              \delta_K(j-l)                            \nonumber  \\
        & = &  4N
              \sigma^4_H.
\end{eqnarray}
Substituting (\ref{Eq:Massive_MIMO_Eq11}) and (\ref{Eq:Massive_MIMO_Eq12}) in
(\ref{Eq:Massive_MIMO_Eq10}) we get
\begin{eqnarray}
\label{Eq:Massive_MIMO_Eq13_2}
\sigma_{U'}^2 = 4 N \sigma^2_H \sigma^2_W +
                8 N (N-1) \sigma^4_H.
\end{eqnarray}
Consider 
\begin{eqnarray}
\label{Eq:Massive_MIMO_Eq14}
\tilde{Y}_i & = & \frac{1}{N_{rt}}
                  \sum_{k=0}^{N_{rt}-1}
                  \tilde{Y}_{k,\, i}               \nonumber  \\
            & = &  F_i S_i +
                  \tilde U_i   \qquad \mbox{for $0 \le i \le N-1$}
\end{eqnarray}
where $\tilde Y_{k,\, i}$ is defined in (\ref{Eq:Massive_MIMO_Eq7}) and
\begin{eqnarray}
\label{Eq:Massive_MIMO_Eq15}
F_i        & = & \frac{1}{N_{rt}}
                 \sum_{k=0}^{N_{rt}-1}
                 \tilde{F}_{k,\, i,\, i}               \nonumber  \\
\tilde U_i & = & \frac{1}{N_{rt}}
                 \sum_{k=0}^{N_{rt}-1}
                 \tilde U_{k,\, i}'.
\end{eqnarray}
Note that $F_i$ in (\ref{Eq:Massive_MIMO_Eq15}) is real-valued. Since
$\tilde U_{k,\, i}'$ is independent over $k$ we have
\begin{eqnarray}
\label{Eq:Massive_MIMO_Eq16}
 E
\left[
\left|
\tilde U_i
\right|^2
\right] & = & \frac{\sigma^2_{U'}}{N_{rt}}           \nonumber  \\
        & = & \frac{1}{N_{rt}}
              \left(
                4 N \sigma^2_H \sigma^2_W +
                8 N (N-1) \sigma^4_H
              \right)                                \nonumber  \\
        & \stackrel{\Delta}{=}
            & \sigma^2_U.
\end{eqnarray}
In other words, the interference plus noise power reduces due to averaging.
The average signal-to-noise ratio per bit in decibels is defined as 
\cite{Vasu_ICWMC2016,Vasu_Adv_Tele_2017} (see also the appendix)
\begin{eqnarray}
\label{Eq:Massive_MIMO_Eq17}
\mbox{SNR}_{\mathrm{av},\, b}
& = &  10
      \log_{10}
      \left(
      \frac{E
      \left[
      \left|
      \sum_{j=0}^{N-1}
      \tilde{H}_{k,\, i,\, j}S_{j}
      \right|^2
      \right]\times 2N_{rt}}
      {E
      \left[
      \left|
      \tilde{W}_{k,\, i}
      \right|^2
      \right]}
      \right)                                             \nonumber  \\
& = &  10
      \log_{10}
      \left(
      \frac{2N\sigma_H^2\times 2\times 2N_{rt}}
           {2\sigma_{W}^{2}}
      \right)                                             \nonumber  \\
& = &  10
      \log_{10}
      \left(
      \frac{4N N_{rt}\sigma_{H}^{2}}
                 {\sigma_{W}^{2}}
      \right).
\end{eqnarray}
From (\ref{Eq:Massive_MIMO_Eq17}) we can write
\begin{equation}
\label{Eq:Massive_MIMO_Eq17_1}
\frac{\sigma^2_W}{N_{rt}} = \frac{4N\sigma^2_H}
                                 {10^{0.1\,\textup{SNR}_{\mathrm{av},\, b}}}.
\end{equation}
Substituting (\ref{Eq:Massive_MIMO_Eq17_1}) in (\ref{Eq:Massive_MIMO_Eq16})
we get
\begin{eqnarray}
\label{Eq:Massive_MIMO_Eq17_2}
\sigma^{2}_{U} & = & \frac{4N\sigma^2_H\sigma^2_W}
                          {N_{rt}} +
                     \frac{8\sigma^{4}_{H}N(N-1)}
                          {N_{rt}}                            \nonumber  \\
               & = & \underbrace{
                     \frac{4N\sigma^2_{H}
                     \cdot
                     \left(
                      4N\sigma^{2}_{H}
                     \right)}
                          {10^{0.1\,\textup{SNR}_{\mathrm{av},\, b}}}}_
                     {
                     \textup{Noise power
                             constant for a
                             given SNR}}                      \nonumber  \\
               &   & \mbox{ } +
                     \underbrace{
                     \frac{8\sigma^{4}_{H}N(N-1)}
                          {N_{rt}}}_
                     {
                     \textup{Interference power
                             reduces with increasing
                             $N_{rt}$}}                       \nonumber  \\
               & = & \frac{16N^2\sigma^4_H}
                          {10^{0.1\textup{SNR}_{\mathrm{av},\, b}}} +
                     \frac{8\sigma^{4}_{H}N(N-1)}
                          {N_{rt}} .
\end{eqnarray}
After concatenation, the signal $\tilde{Y}_i$ and $F_{i,\, i}$ in
(\ref{Eq:Massive_MIMO_Eq14}) for $0\leq i\leq L_{d}-1$ is sent to the turbo
decoder \cite{Vasu_Book10}, as explained below.
\subsection{Turbo Decoding - the BCJR Algorithm}
\label{SSec:BCJR}
\begin{figure}[tbh]
\begin{center}
\input{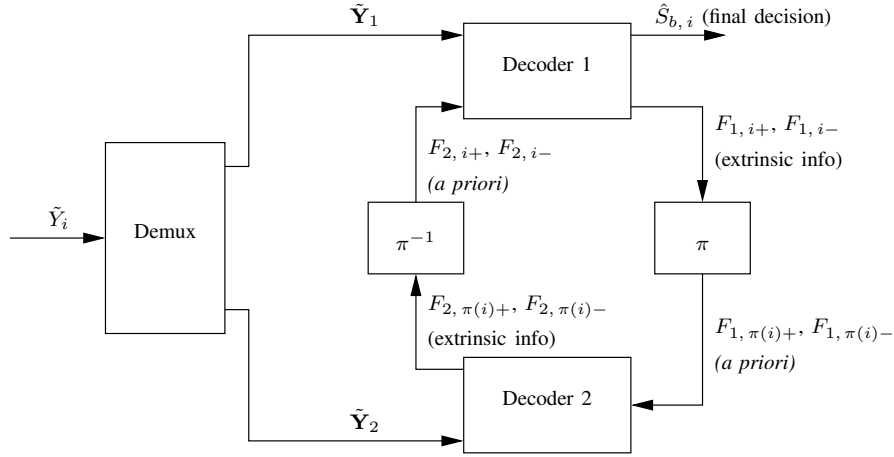}
\caption{Turbo decoder.}
\label{Fig:Turbo_Dec1}
\end{center}
\end{figure}
The block diagram of the turbo decoder is depicted in 
Figure~\ref{Fig:Turbo_Dec1}. Note that
\begin{equation}
\label{Eq:Massive_MIMO_Eq17_2_1}
\begin{aligned}
\tilde{\mathbf{Y}}_1 & = \left[
                         \begin{array}{ccc}
                         \tilde Y_0         & \ldots & \tilde Y_{L_{d1}-1}
                         \end{array}
                         \right]       \\
\tilde{\mathbf{Y}}_2 & = \left[
                         \begin{array}{ccc}
                         \tilde Y_{L_{d1}}  & \ldots & \tilde Y_{L_{d}-1}
                         \end{array}
                         \right].
\end{aligned}
\end{equation}
The BCJR algorithm has the following components:
\begin{enumerate}
 \item The forward recursion
 \item The backward recursion
 \item The computation of the extrinsic information and the final
       {\it a posteriori\/} probabilities.
\end{enumerate}
Let $\mathscr{S}$ denote the number of states in the encoder trellis. Let
$\mathscr{D}_n$ denote the set of states that diverge from state $n$,
for $0 \le n \le \mathscr{S} -1$. For example
\begin{equation}
\label{Eq:Massive_MIMO_Eq17_3}
\mathscr{D}_0 = \left
                \{
                 0,\, 3
                \right
                \}
\end{equation}
implies that states 0 and 3 can be reached from state 0. Similarly, let
$\mathscr{C}_n$ denote the set of states that converge to state $n$.
Let $\alpha_{i,\, n}$ denote the forward sum-of-products (SOP) at time
$i$ ($\left(0\leq i\leq L_{d1}-2\right)$) at state $n$. Then the forward SOP
for decoder 1 can be recursively computed as follows (forward recursion)
\cite{Vasu_Book10}
\begin{eqnarray}
\label{Eq:Massive_MIMO_Eq17_4}
\begin{aligned}
\alpha_{i+1,\, n}' & =   \sum_{m\in \mathscr{C}_n}
                         \alpha_{i,\, m}
                         \gamma_{1,\, i,\, m,\, n}
                          P
                         \left(
                          S_{b,\, i,\, m,\, n}
                         \right)                                            \\
\alpha_{0,\, n}    & =    1
                         \qquad
                         \mbox{for $0 \le n \le \mathscr{S}-1$}             \\
\alpha_{i+1,\, n}  & =   \alpha_{i+1,\, n}'
                         \Bigg /
                         \left(
                         \sum_{n=0}^{\mathscr{S}-1}
                         \alpha_{i+1,\, n}'
                         \right)
\end{aligned}
\end{eqnarray}
where
\begin{equation}
\label{Eq:Massive_MIMO_Eq17_5}
 P
\left(
 S_{b,\, i,\, m,\, n}
\right) =
\left\{
\begin{array}{ll}
F_{2,\, i+} & \mbox{if $S_{b,\, i,\, m,\, n} = +1$}\\ 
F_{2,\, i-} & \mbox{if $S_{b,\, i,\, m,\, n} = -1$}
\end{array}
\right.
\end{equation}
denotes the {\it a priori\/} probability of the systematic (data) bit
corresponding to the transition from state $m$ to state $n$, at decoder 1 at
time $i$, obtained from the $2^{nd}$ decoder at time $l$, after de-interleaving,
that is, $i=\pi^{-1}(l)$ for some $0 \leq l\leq L_{d1}-1$, $l\neq i$ and
\begin{eqnarray}
\label{Eq:Massive_MIMO_Eq18}
\gamma_{1,\, i,\, m,\, n} = \exp
                            \left[
                             -
                            \,
                            \frac{
                            \left|
                            \tilde Y_i - F_i S_{m,\, n}
                            \right|^2}{2\sigma^2_U}
                            \right]
\end{eqnarray}
where $S_{m,\, n}$ denotes the coded QPSK symbol corresponding to the transition
from state $m$ to $n$ in the trellis. The normalization step in the last 
equation of (\ref{Eq:Massive_MIMO_Eq17_4}) is done
to prevent numerical instabilities.

Let $\beta_{i,\, n}$ denote the backward SOP at time $i$
$\left(1\leq i\leq L_{d1}-1\right)$ at state $n$
$\left(0\leq n\leq \mathscr{S}-1 \right)$. Then the recursion for the backward
SOP (backward recursion) at decoder 1 can be written as:
\begin{equation}
\label{Eq:Massive_MIMO_Eq18_1}
\begin{aligned} 
\beta_{i,\, n}' & = \sum_{m \in \mathscr{D}_n}
                    \beta_{i+1,\, m}
                    \gamma_{1,\, i,\, n,\, m}
                     P
                    \left(
                     S_{b,\, i,\, n,\, m}
                    \right)                                            \\
\beta_{L_{d1},\, n}
                & =  1
                    \qquad
                    \mbox{for $0\leq n\leq \mathscr{S}-1$}             \\
\beta_{i,\, n}  & = \beta_{i,\, n}'
                    \Bigg /
                    \left(
                    \sum_{n=0}^{\mathscr{S}-1}
                    \beta_{i,\, n}'
                    \right).
\end{aligned}
\end{equation}
Once again, the normalization step in the last equation of 
(\ref{Eq:Massive_MIMO_Eq18_1}) is done to prevent numerical instabilities.

Let $\rho^{+}(n)$ denote the state that is reached from state $n$ when the
input symbol is $+1$. Similarly let $\rho^{-}(n)$ denote the state that can be
reached from state $n$ when the input symbol is $-1$. Then the extrinsic
information from decoder 1 to 2 is calculated as follows for
$0 \le i \le L_{d1}-1$
\begin{equation}
\label{Eq:Massive_MIMO_Eq21}
\begin{aligned}
G_{\mathrm{norm},\, i+} & = \sum_{n=0}^{\mathscr{S}-1}
                            \alpha_{i,\, n}
                            \gamma_{1,\, i,\, n,\,\rho^{+}(n)}
                            \beta_{i+1,\, \rho^{+}(n)}              \\
G_{\mathrm{norm},\, i-} & = \sum_{n=0}^{\mathscr{S}-1}
                            \alpha_{i,\, n}
                            \gamma_{1,\, i,\, n,\,\rho^{-}(n)}
                            \beta _{i+1,\,\rho^{-}(n)}
\end{aligned}
\end{equation}
which is further normalized to obtain
\begin{equation}
\label{Eq:Massive_MIMO_Eq22}
\begin{aligned}
F_{1,\, i+} & =  G_{\mathrm{norm},\, i+}
                \big /
                \left(
                 G_{\mathrm{norm},\, i+}+G_{\textup{norm},\, i-}
                \right)                      \\
F_{1,\, i-} & =  G_{\mathrm{norm},\, i-}
                \big /
                \left(
                 G_{\mathrm{norm},\, i+}+G_{\textup{norm},\, i-}
                \right).
\end{aligned}
\end{equation}
Equations (\ref{Eq:Massive_MIMO_Eq17_4}), (\ref{Eq:Massive_MIMO_Eq18_1}),
(\ref{Eq:Massive_MIMO_Eq21}) and (\ref{Eq:Massive_MIMO_Eq22}) constitute the
MAP recursions for the first decoder. The MAP recursions for the second
decoder are similar excepting that $\gamma_{1,\, i,\, m,\, n}$ is replaced by
\begin{eqnarray}
\label{Eq:Massive_MIMO_Eq19}
\gamma_{2,\, i,\, m,\, n} = \exp
                            \left[
                             -
                            \,
                            \frac{
                            \left|
                            \tilde Y_{i1} - F_{i1} S_{m,\, n}
                            \right|^2}{2\sigma^2_U}
                            \right]
\end{eqnarray}
where $\tilde Y_{i1}$ and $F_{i1}$ are obtained by concatenating
$\tilde Y_i$ and $F_{i,\, i}$ in (\ref{Eq:Massive_MIMO_Eq14}) and
\begin{equation}
\label{Eq:Massive_MIMO_Eq20}
i1 = i + L_{d1}  \qquad \mbox{for $0 \le i \le L_{d1}-1$}
\end{equation}
and $F_{1,\, i+}$, $F_{1,\, i-}$ in (\ref{Eq:Massive_MIMO_Eq22}) is
replaced by $F_{2,\, i+}$ and $F_{2,\, i-}$ respectively (see 
Figure~\ref{Fig:Turbo_Dec1}).

After several iterations, the final {\it a posteriori\/} probabilities of the
$i^{th}$ data bit obtained at the output of the first decoder is computed as
(for $0 \le i \le L_{d1}-1$):
\begin{equation}
\label{Eq:Massive_MIMO_Eq23}
\begin{aligned} 
 P
\left(
 S_{b,\, i} = +1|\tilde{\mathbf{Y}}_1,\,\tilde{\mathbf{Y}}_2
\right) & = G_{\mathrm{norm},\, i+} F_{2,\, i+}       \\
 P
\left(
 S_{b,\, i} = -1|\tilde{\mathbf{Y}}_1,\,\tilde{\mathbf{Y}}_2
\right) & = G_{\mathrm{norm},\, i-} F_{2,\, i-}
\end{aligned}
\end{equation}
where again $F_{2,\, k+}$ and $F_{2,\, k-}$ denote the {\it a priori\/} 
probabilities obtained at the output of the second decoder (after
de-interleaving) in the previous iteration. The final estimate of the
$i^{th}$ data bit is given as (see Figure~\ref{Fig:Turbo_Dec1}):
\begin{equation}
\label{Eq:Massive_MIMO_Eq23_1}
\begin{array}{ll} 
\mbox{Choose $\hat S_{b,\, i}=+1$ if} &
 P
\left(
 S_{b,\, i} = +1|\tilde{\mathbf{Y}}_1,\,\tilde{\mathbf{Y}}_2
\right)
>
 P
\left(
 S_{b,\, i} = -1|\tilde{\mathbf{Y}}_1,\,\tilde{\mathbf{Y}}_2
\right)
\\
\mbox{Choose $\hat S_{b,\, i}=-1$ if} &
 P
\left(
 S_{b,\, i} = -1|\tilde{\mathbf{Y}}_1,\,\tilde{\mathbf{Y}}_2
\right)
>
 P
\left(
 S_{b,\, i} = +1|\tilde{\mathbf{Y}}_1,\,\tilde{\mathbf{Y}}_2
\right).
\end{array}
\end{equation}
Note that:
\begin{enumerate}
 \item One iteration involves decoder 1 followed by decoder 2.
 \item Since the terms $\alpha_{i,\, n}$ and $\beta_{i,\, n}$ depend on
       $F_{2,\, i+}$, $F_{2,\, i-}$ for decoder 1, and $F_{1,\, i+}$,
       $F_{1,\, i-}$ for decoder 2, they have to be recomputed for every
       decoder in every iteration according to (\ref{Eq:Massive_MIMO_Eq17_4})
       and (\ref{Eq:Massive_MIMO_Eq18_1}) respectively.
\end{enumerate}
In the computer simulations, robust turbo decoding \cite{Vasudevan2015} has
been incorporated, that is, the exponent in (\ref{Eq:Massive_MIMO_Eq18}) and
(\ref{Eq:Massive_MIMO_Eq19}) is normalized to the range $[-30,\, 0]$.
\section{Simulation Results and Discussion}
\label{Sec:Sim_Results}
In this section, we present the results from computer simulations. The
simulation parameters are presented in Table~\ref{Tbl:Sim_Param}.
\begin{table}[tbh]
\begin{center}
\caption{Simulation parameters.}
\input{sim_param.pstex_t}
\label{Tbl:Sim_Param}
\end{center}
\end{table}
At high SNR, the number of frames simulated is $10^6$, whereas for low and
medium SNR, the number of frames simulated is $10^5$.
\begin{figure}[tbh]
\begin{center}
\input{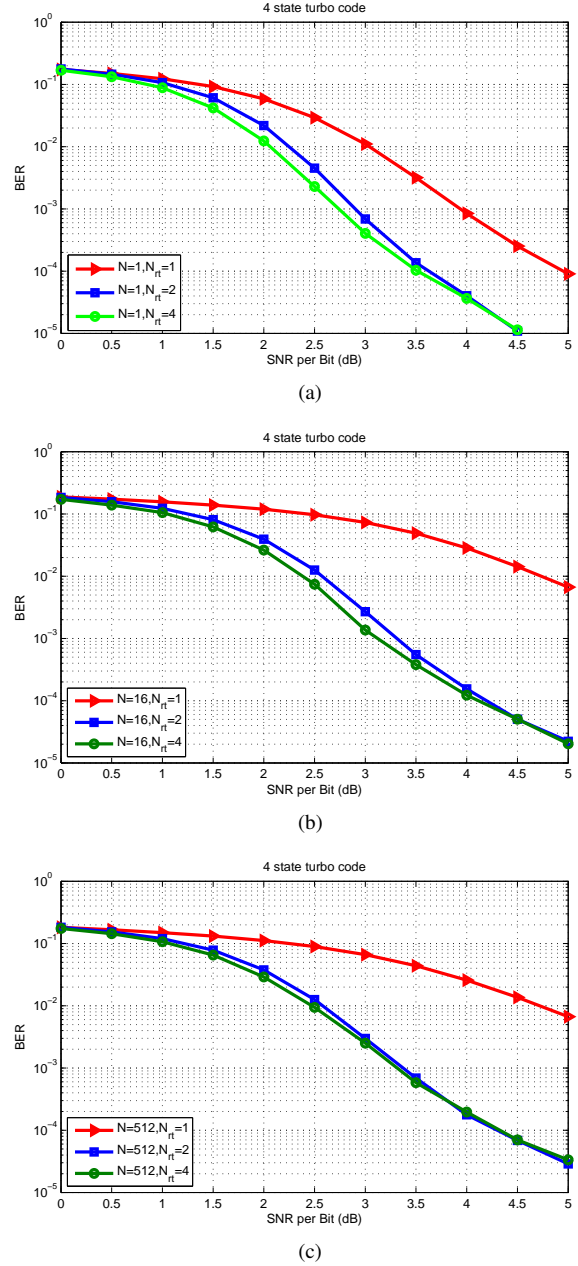}
\caption{Results for the 4-state turbo code in (\ref{Eq:Massive_MIMO_Eq24}).}
\label{Fig:MMIMO_4St}
\end{center}
\end{figure}
In Figure~\ref{Fig:MMIMO_4St}, we present the simulation results for a 
4-state turbo code with generating matrix given by
\begin{equation}
\label{Eq:Massive_MIMO_Eq24}
\mathbf{G}(D) = \left[
                \begin{array}{cc}
                 1 & \tfrac{1+D^{2}}{1+D+D^{2}}
                \end{array}
                \right].
\end{equation}
From Figures~\ref{Fig:MMIMO_4St}(a) to (c) we see the following.
\begin{enumerate}
 \item There is no significant degradation in the BER performance due to the
       increase in the number of antennas ($N$), for a given number of
       re-transmissions $N_{rt}>1$. For example, with $N_{rt}=2$ and $N=16$,
       a BER of $10^{-4}$ is attained at an SNR per bit of 4dB, whereas the
       same BER is attained at an SNR per bit of 4.25 dB for $N=512$ -- this
       is just a 0.25 dB degradation in performance. Observe that the spectral
       efficiency with $N=16$ antennas and $N_{rt}=2$ re-transmissions, is
       4 bits/transmission or 4 bits/sec/Hz, since each QPSK symbol carries
       $1/4$ bits of information (see appendix). However, the spectral
       efficiency with $N=512$ antennas and $N_{rt}=2$ re-transmissions is 128
       bits/sec/Hz. In other words,
       an increase in the spectral efficiency by a factor of 32 results in
       only a 0.25 dB degradation in the BER performance.
 \item With $N_{rt}=2$, there is significant improvement in BER performance
       compared to $N_{rt}=1$, for all values of $N$. However the BER
       performance with $N_{rt}=4$ is comparable to $N_{rt}=2$. This is
       because, with increasing $N_{rt}$ the BER is limited by the variance of
       the noise term in (\ref{Eq:Massive_MIMO_Eq17_2}), even though the
       variance of the interference term gets reduced due to averaging.
 \item Note that when $N=1$, the interference is zero and only noise is present.
       We see from Figure~\ref{Fig:MMIMO_4St}(a) that there is a significant
       improvement in performance for $N_{rt}=2$, compared to $N_{rt}=1$. This
       can be attributed to the fact that $F_i$ in (\ref{Eq:Massive_MIMO_Eq15})
       contains two positive terms (independent Rayleigh distributed random
       variables) for $N_{rt}=2$ compared to $N_{rt}=1$. Hence, the probability
       that both terms are simultaneously close to zero, is small.
 \item It is interesting to compare the case $N=N_{rt}=1$ in
       Figure~\ref{Fig:MMIMO_4St}(a) with Figure~12 in
       \cite{Vasu_Adv_Tele_2017} with $N_r=1$. Both systems are identical,
       in terms of the received signal model, that is
\begin{equation}
\label{Eq:Massive_MIMO_Eq24_1}
\tilde R_i = \tilde H_i S_i + \tilde W_i
\end{equation}
       where $i$ denotes the time index. In this work, we obtain a BER of
       $10^{-4}$ at an average SNR per bit of 5 dB, whereas in
       \cite{Vasu_Adv_Tele_2017} we obtain the same BER at an average SNR per
       bit of just 2.25 dB. What could be the reason for this difference? The
       answer lies in the computation of gammas. In this
       work, the gammas are computed using (\ref{Eq:Massive_MIMO_Eq18}) and
       (\ref{Eq:Massive_MIMO_Eq19}), which is sub-optimum compared to (66)
       in \cite{Vasu_Adv_Tele_2017}. This is because, the noise term
       $\tilde U_i$ in (\ref{Eq:Massive_MIMO_Eq14}) is equal to
       $\tilde H_i^*\tilde W_i$, which is not even Gaussian (recall that
       $\tilde U_i$ is Gaussian for large values of $N$ due to the central
       limit theorem). However, in this work, we are assuming that $\tilde U_i$
       is Gaussian, for $N=1$.
\end{enumerate}
\begin{figure}[tbh]
\begin{center}
\input{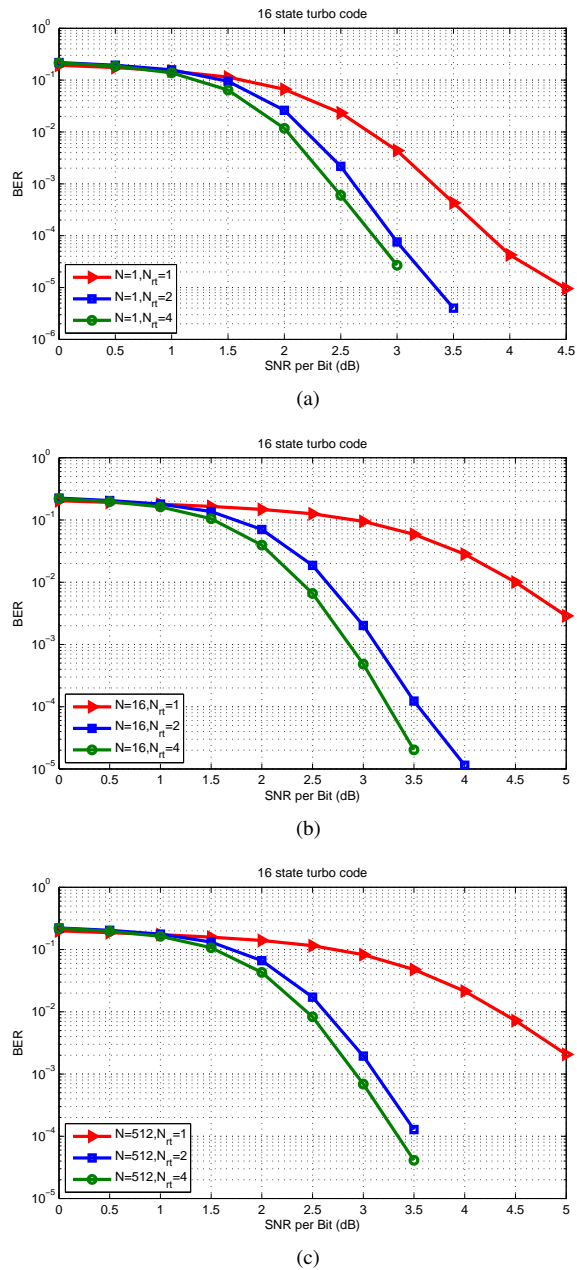}
\caption{Results for the 16-state turbo code in (\ref{Eq:Massive_MIMO_Eq25}).}
\label{Fig:MMIMO_16St}
\end{center}
\end{figure}
In Figure~\ref{Fig:MMIMO_16St}, we present the simulation results for a 
16-state turbo code with generating matrix given by \cite{Vasu_Book10}
\begin{equation}
\label{Eq:Massive_MIMO_Eq25}
\mathbf{G}(D) = \left[
                \begin{array}{cc}
                 1 & \frac{1+D^{2}+D^{3}+D^{4}}{1+D+D^{4}}
                \end{array}
                \right].
\end{equation}
We observe the following in Figures~\ref{Fig:MMIMO_16St}(a)-(c):
\begin{enumerate}
 \item There is again a significant improvement in BER performance for
       $N_{rt}=2$, compared to $N_{rt}=1$. However, the improvement in BER
       for $N_{rt}=4$ is not much, compared to $N_{rt}=2$.
 \item Comparing Figures~\ref{Fig:MMIMO_4St} and \ref{Fig:MMIMO_16St}, with
       $N=16$ and $N_{rt}=2$, the encoder in (\ref{Eq:Massive_MIMO_Eq25})
       gives only a 0.5 dB improvement at a BER of $10^{-4}$, over the encoder
       in (\ref{Eq:Massive_MIMO_Eq24}).
 \item Comparing Figures~\ref{Fig:MMIMO_4St} and \ref{Fig:MMIMO_16St}, with
       $N=512$ and $N_{rt}=2$, the encoder in (\ref{Eq:Massive_MIMO_Eq25})
       gives only a 0.75 dB improvement at a BER of $10^{-4}$, over the encoder
       in (\ref{Eq:Massive_MIMO_Eq24}). These results indicate that this may
       not be the best 16-state turbo code.
\end{enumerate}
\section{Conclusion}
\label{Sec:Conclude}
We have shown by analysis as well as computer simulations that, as the number of
retransmissions increase, the BER decreases. There is little improvement by
using a 16-state turbo code as compared to the 4-state code, in terms of the
BER. Perhaps, this may not be the best 16-state turbo code. Future work could
be to use iterative interference cancellation with no re-transmissions, since
the re-transmissions reduce the spectral efficiency. Estimating the $N\times N$
channel matrix is also a good topic for future research.









\section{appendix}
\label{Sec:Appendix}
We derive the minimum average SNR per bit required for error-free propagation
over a massive MIMO channel with re-transmissions. Consider the signal
\begin{equation}
\label{Eq:Massive_MIMO_Eq25_1}
\tilde r_i = \tilde x_i + \tilde w_i
\end{equation}
where the subscript $i$ denotes the time index, $\tilde x_i$ is the transmitted
signal (message) and $\tilde w_i$ denotes samples of zero-mean noise, not
necessarily Gaussian, with variance per dimension equal to $\sigma^2_w$. All
the terms in (\ref{Eq:Massive_MIMO_Eq25_1}) are complex-valued or
two-dimensional. Here the
term ``dimension'' refers to a communication link between the
transmitter and the receiver carrying only real-valued signals
\cite{Vasu_ICWMC2016,Vasu_Adv_Tele_2017}.
The number of bits per transmission, defined as the channel capacity, is given
by \cite{Salehi_Dig,Vasu_ICWMC2016,Vasu_Adv_Tele_2017}
\begin{equation}
\label{Eq:Massive_MIMO_Eq26}
C = \log_2
    \left(
     1+\textup{SNR}
    \right)
    \qquad \textup{bits per transmission}
\end{equation}
over a complex dimension, where the average SNR is given by
\begin{eqnarray}
\label{Eq:Massive_MIMO_Eq27}
\textup{SNR} & = & \frac{
                    E
                   \left[
                   \left|
                   \tilde x_i
                   \right|^2
                   \right]}
                          {
                    E
                   \left[
                   \left|
                   \tilde w_i
                   \right|^2
                   \right]}      \nonumber  \\
             & = & \frac{P_{\mathrm{av}}'}{2\sigma^2_w}
\end{eqnarray}
over a complex dimension. Recall that (\ref{Eq:Massive_MIMO_Eq26}) gives the
minimum SNR for the error-free propagation of $C$ bits.
\begin{Propo}
\label{Prop:Prop1}
The channel capacity is additive over the number of complex dimensions. In other
words, the channel capacity over $N$ complex dimensions, is equal to the sum of
the capacities over each complex dimension, provided the information is
independent across the complex dimensions
\cite{Vasudevan2015,Vasu_ICWMC2016,Vasu_Adv_Tele_2017}. Independence of
information also implies that, the bits transmitted over one complex dimension
is not the interleaved version of the bits transmitted over any
other complex dimension.
\end{Propo}
\begin{Propo}
\label{Prop:Prop2}
Conversely, if $C$ bits per transmission are sent over $N$ complex dimensions,
it seems reasonable to assume that each complex dimension receives
$C/N$ bits per transmission
\cite{Vasudevan2015,Vasu_ICWMC2016,Vasu_Adv_Tele_2017} .
\end{Propo}
The reasoning for Proposition~\ref{Prop:Prop2} is as follows. We assume
that a ``bit'' denotes ``information''. Now, if each of the $N$
antennas (complex dimensions) receive the ``same'' $C$ bits of
information, then we might as well have only one antenna,
since the other antennas are not yielding any additional information.
On the other hand, if each of the $N$ antennas receive
``different'' $C$ bits of information, then we end up receiving
more information ($CN$ bits) than what we transmit ($C$ bits),
which is not possible. Therefore, we assume that each complex
dimension receives $C/N$ bits of ``different'' information.

Observe that the average SNR in (\ref{Eq:Massive_MIMO_Eq27}) is not the 
average SNR per bit over a complex dimension. In order to compute the average
SNR per bit, we note from Figure~\ref{Fig:System_Model} that each data
bit generates two QPSK symbols, and each QPSK symbol is repeated $N_{rt}$ times.
Therefore, from Proposition~\ref{Prop:Prop2}, each QPSK symbol carries
$1/(2N_{rt})$ bits of information. The information sent in one transmission
is $N/(2N_{rt})$ bits, from the $N$ transmit antennas
(Proposition~\ref{Prop:Prop1}). The information in each receive
antenna in one transmission over a complex dimension is (Proposition~\ref{Prop:Prop2}):
\begin{eqnarray}
\label{Eq:Massive_MIMO_Eq27_1}
N/(2NN_{rt}) = 1/(2N_{rt}) = C \qquad \mbox{bits}
\end{eqnarray}
which is identical to the channel capacity in (\ref{Eq:Massive_MIMO_Eq26}).

Let us now consider the $i^{th}$ element of $\tilde{\mathbf{R}}_k$ in
(\ref{Eq:Massive_MIMO_Eq1}). We have
\begin{eqnarray}
\label{Eq:Massive_MIMO_Eq28}
\tilde R_{k,\, i} = \sum_{j=1}^{N}
                    \tilde H_{k,\, i,\, j}
                     S_j +
                    \tilde W_{k,\, i}.
\end{eqnarray}
Now, if we substitute
\begin{eqnarray}
\label{Eq:Massive_MIMO_Eq29}
\tilde x_i & = & \sum_{j=1}^{N}
                 \tilde H_{k,\, i,\, j}
                  S_j                               \nonumber  \\
\tilde w_i & = & \tilde W_{k,\, i}
\end{eqnarray}
in (\ref{Eq:Massive_MIMO_Eq25_1}), the channel capacity remains unchanged,
as given in (\ref{Eq:Massive_MIMO_Eq26}), with SNR equal to
\begin{eqnarray}
\label{Eq:Massive_MIMO_Eq30}
\textup{SNR} = \frac{2NP_{\mathrm{av}}\sigma^2_H}{2\sigma^2_W}
\end{eqnarray}
where $\sigma^2_H$, $\sigma^2_W$ and $P_{\mathrm{av}}$ are defined in
(\ref{Eq:Massive_MIMO_Eq2}), (\ref{Eq:Massive_MIMO_Eq3}) and
(\ref{Eq:Massive_MIMO_Eq13}) respectively. However, the information contained
in $\tilde R_{k,\,i}$ in (\ref{Eq:Massive_MIMO_Eq28}) is $1/(2N_{rt})$ bits
(see (\ref{Eq:Massive_MIMO_Eq27_1})), hence the SNR in
(\ref{Eq:Massive_MIMO_Eq30}) is for $1/(2N_{rt})$ bits.
Therefore, the SNR per bit is
\begin{eqnarray}
\label{Eq:Massive_MIMO_Eq31}
\textup{SNR}_{\mathrm{av},\, b}
& = & \frac{2NP_{\mathrm{av}}\sigma^2_H\cdot 2N_{rt}}
           {2\sigma^2_W}      \nonumber  \\
& = & \frac{\textup{SNR}}{C}
\end{eqnarray}
where we have used (\ref{Eq:Massive_MIMO_Eq27_1}). Substituting
(\ref{Eq:Massive_MIMO_Eq31}) in (\ref{Eq:Massive_MIMO_Eq26}) we get
\begin{eqnarray}
\label{Eq:Massive_MIMO_Eq33}
C   =   \log_2
        \left(
         1 +
         C
        \,
        \textup{SNR}_{\mathrm{av},\, b}
        \right)
        \qquad
        \textup{bits per transmission}
\end{eqnarray}
over a complex dimension. Re-arranging terms in (\ref{Eq:Massive_MIMO_Eq33})
we get
\begin{eqnarray}
\label{Eq:Massive_MIMO_Eq34}
\textup{SNR}_{\mathrm{av},\, b} = \frac{2^C - 1}{C}.
\end{eqnarray}
Thus (\ref{Eq:Massive_MIMO_Eq34}) implies that as $C\rightarrow 0$,
$\textup{SNR}_{\mathrm{av},\, b}\rightarrow \ln(2)\equiv -1.6$ dB, which is the
minimum average SNR per bit required for error-free propagation over a
massive MIMO channel, with re-transmissions. Just as in the case of turbo
codes, it may not be necessary for $C$ to approach zero, in order to attain
the channel capacity.
\bibliographystyle{spphys}
\bibliography{mybib,mybib1,mybib2,mybib3,mybib4}
\end{document}